\documentclass[prd,aps,floatfix,nofootinbib,11 pt]{revtex4}
\usepackage{amsmath,graphicx,color,epsfig}

\input{tcilatex}

\begin{document}

\title{Multiparty quantum cryptographic protocol}
\author{M. Ramzan\thanks{%
mramzan@phys.qau.edu.pk} and M. K. Khan}
\address{Department of Physics Quaid-i-Azam University \\
Islamabad 45320, Pakistan}

\begin{abstract}
We propose a multiparty quantum cryptographic protocol. Unitary operators
applied by Bob and Charlie, on their respective qubits of a tripartite
entangled state encodes a classical symbol that can be decoded at Alice's
end with the help of a decoding matrix. Eve's presence can be detected by
the disturbance of the decoding matrix. Our protocol is secure against
intercept-resend attacks. Furthermore, it is efficient and deterministic in
the sense that two classical bits can be transferred per entangled pair of
qubits. It is worth mentioning that in this protocol same symbol can be used
for key distribution and Eve's detection that enhances the efficiency of the
protocol.\newline
\end{abstract}

\pacs{03.67.-a; 03.67.Hk; 03.67.Dd}
\maketitle

\address{Department of Physics Quaid-i-Azam University \\
Islamabad 45320, Pakistan}

Keywords: Quantum cryptography; Eve's detection; Decoding matrix; Secure
communication\newline

\vspace*{1.0cm}

\vspace*{1.0cm}



\section{Introduction}

Quantum key distribution (QKD) provides a secure way for generating a
private key between two remote parties and then it can be used to distribute
the key among several parties. In public key crypto-systems, such as, Rivest
Shamir Adleman (RSA), the receiver generates a pair of keys: a public key
and a private key \cite{Rivest}. The security of the communication relies on
determining the prime factors of a public key, a large integer. The public
key is used to encrypt the message while the private key decrypts it. With
the advent of quantum computing, it is now possible to factorize very large
numbers much faster. As a result the security of the RSA can easily be
compromised. The first key distribution protocol using four quantum states
was proposed by Bennett and Brassard \cite{BB84}. In 1991, Artur Ekert \cite%
{Ekert}, by using a different approach to quantum cryptography, proposed a
key distribution protocol in which entangled pairs of qubits are distributed
to Alice and Bob, who then extract key bits by measuring their qubits.
Enormous efforts have been made to develop cryptographic protocols based on
quantum mechanics [4-11].

Quantum key distribution is a process in which the legitimate users of
communication first establish a shared secret key by transmitting a
classical message and then using that key to encrypt (decrypt) the secret
message. In the field of quantum cryptography or quantum key distribution,
important advances have been made in both theoretical and experimental
directions. However, many open problems remain in both fields, such as
bringing theory and application together. Indeed, the theoretical tools have
recently been applied to study the security of practical implementations 
\cite{Gott}. Furthermore, QKD has been shown to be experimentally feasible 
\cite{Honjo}.

In recent years researchers have drawn attention to QKD protocols that
involve multilevel systems with two parties [14-17], or multiple parties
with two-level systems \cite{Durt}. Motivation of multilevel quantum key
distribution is that more information can be carried that may increase the
information flux. Some multilevel protocols have been shown to have greater
security against eavesdropping attacks \cite{Durt,Burb}. Quantum
cryptography offers an entirely new technique for secure key distribution
where security relies upon the laws of quantum physics instead of
computational complexity. There are several protocols for quantum
cryptography [18-23] and quantum secure direct communication [24-26]\ which
involve three-party communication. The \textquotedblleft quantum
dialogue\textquotedblright\ protocol proposed in reference \cite{Ba}, offers
a direct way to exchange confidential messages, defeats the disturbance
attack but, in turn, it is vulnerable to the intercept-resend attack \cite%
{Xia}.

In this paper, we propose a three-party (Alice, Bob and Charlie) quantum
cryptographic protocol by using Greenberger-Horne-Zeilinger (GHZ) triplet
entangled states. In our protocol, Alice can obtain the secret messages of
Bob and Charlie by decoding the secrets sent by them. Their secret messages
exchange is secure and simultaneous. Therefore, our protocol may be feasible
in near-future technology. Our protocol is secure and efficient in the sense
that same symbol can be used for key distribution and Eve's detection that
enhances the efficiency of the protocol.

\section{The protocol}

We consider a multiparty cryptographic task where three parties, Alice, Bob,
and Charlie, would like to obtain a random string of numbers, so that it can
be called as a quantum network. In our protocol, the three users of
communication Alice, Bob and Charlie share a large number of entangled
tripartite GHZ states (see figure 1). Alice prepares entangled GHZ triplet
states, and sends the second qubit to Bob and third qubit to Charlie
respectively, who then apply their local operations on their individual
qubits and send it back to Alice.\textrm{\ }Alice first applies her local
operators on her part of qubit and then she performs GHZ measurements. Prior
to any key distribution Alice (knowing the set of operators Bob and Charlie
would apply), simulate their actions and construct a decoding matrix with
elements as expectation values of the coding operators for herself, Bob and
Charlie (see table 1). Once this decoding matrix is constructed, Alice
applies her local unitary operators and measure the expectation values for
the decoding operators. The elements of the decoding matrix depend on the
operators applied by Alice, Bob and Charlie. Upon comparing this expectation
value against the already constructed decoding matrix, Alice would be able
to identify the unitary operators applied by Bob and Charlie. If Eve
performs a measurement on the qubit during transmission, the expectation
value recorded by Alice would be different and she would be able to detect
Eve's presence and will use the classical channel to inform Bob and Charlie
to ignore that particular key. By repeating this process for several times a
string of bits is transferred from Bob and Charlie to Alice which acts as a
secret key for communication.

Whenever an eavesdropper, Eve, makes any type of measurement in the way then
the quantum state is disturbed which results a change in the quantum
decoding matrix that helps Alice to detect the presence of eavesdropper.\ It
is worth to note that in this protocol, Alice needs not to have a different
record of values for the detection of eavesdropper as in the case of ref. 
\cite{Ekert}. In our protocol, whenever an eavesdropper, Eve, interferes she
can very easily be detected by the disturbance of the decoding matrix. We
check for the security of the protocol for intercept/re-send attacks and
find it secure against such type of attacks.

Let's consider that the three parties Alice, Bob and Charlie share a
sequence of GHZ triplets in the state%
\begin{equation}
\left\vert \psi _{ABC}\right\rangle =\frac{1}{\sqrt{2}}\left( \left\vert
000\right\rangle +i\left\vert 111\right\rangle \right) _{ABC}
\end{equation}%
Now the three parties can locally manipulate their individual qubits. The
local operators of Alice, Bob and Charlie can be represented by the unitary
operator $U_{i}$\ of the form \cite{Ramzan}%
\begin{equation}
U_{i}=\cos \frac{\theta _{i}}{2}R_{i}+\sin \frac{\theta _{i}}{2}P_{i}
\end{equation}%
where $i=A,$\ $B$ and $C$\ and $R_{i}$, $P_{i}$\emph{\ }are the local
operators defined as:%
\begin{align}
R_{A}\left\vert 0\right\rangle & =e^{i\alpha _{A}}\left\vert 0\right\rangle ,%
\text{\qquad \qquad }R_{A}\left\vert 1\right\rangle =e^{-i\alpha
_{A}}\left\vert 1\right\rangle ,  \notag \\
P_{A}\left\vert 0\right\rangle & =e^{i\left( \frac{\pi }{2}-\beta
_{A}\right) }\left\vert 1\right\rangle ,\text{\qquad }P_{A}\left\vert
1\right\rangle =e^{i\left( \frac{\pi }{2}+\beta _{A}\right) }\left\vert
0\right\rangle ,  \notag \\
R_{B}\left\vert 0\right\rangle & =\left\vert 0\right\rangle ,\text{\qquad
\qquad \qquad }R_{B}\left\vert 1\right\rangle =\left\vert 1\right\rangle , 
\notag \\
P_{B}\left\vert 0\right\rangle & =\left\vert 1\right\rangle ,\text{\qquad
\qquad \qquad }P_{B}\left\vert 1\right\rangle =-\left\vert 0\right\rangle , 
\notag \\
R_{C}\left\vert 0\right\rangle & =\left\vert 0\right\rangle ,\text{\qquad
\qquad \qquad }R_{C}\left\vert 1\right\rangle =\left\vert 1\right\rangle , 
\notag \\
P_{C}\left\vert 0\right\rangle & =\left\vert 1\right\rangle ,\text{\qquad
\qquad \qquad }P_{C}\left\vert 1\right\rangle =-\left\vert 0\right\rangle ,
\end{align}%
where $0\leq \theta _{i}\leq \pi ,$ $-\pi \leq \{\alpha _{A},$ $\beta
_{A}\}\leq \pi .$\emph{\ }Let Alice, Bob and Charlie agree on that Alice can
perform the unitary operators $U_{A}\left( \theta _{A},\alpha _{A},\beta
_{A}\right) $, while Bob and Charlie, can apply the\emph{\ }unitary
operators $U_{B}\left( \theta _{B}\right) $\ and $U_{C}\left( \theta
_{C}\right) $ respectively. Let all the three parties decide prior to any
communication that Alice can apply unitary operators $U_{A}\left(
0,0,0\right) $ and $U_{A}\left( \pi ,\pi ,\pi \right) $. Whereas Bob can
encode one bit of classical information by the two unitary operations as $%
U_{B}\left( 0\right) \rightarrow m_{1},$ and $U_{B}\left( \pi \right)
\rightarrow m_{2}$. On the other hand, Charlie can encode one bit of
classical information by his two unitary operations as $U_{C}\left( 0\right)
\rightarrow m_{3},$ and $U_{C}\left( \pi \right) \rightarrow m_{4}$.\ With
the application of local operations of the communicating parties, the
initial state transforms as 
\begin{equation}
\rho _{f}=(U_{A}\otimes U_{B}\otimes U_{C})\rho _{ABC}(U_{A}\otimes
U_{B}\otimes U_{C})^{\dagger }
\end{equation}%
where $\rho _{ABC}=\left\vert \psi _{ABC}\right\rangle \left\langle \psi
_{ABC}\right\vert $ is the density matrix for the quantum state with the
basis ordered as $\left\vert 000\right\rangle ,$ $\left\vert
001\right\rangle ,$ $\left\vert 100\right\rangle ,$ $\left\vert
101\right\rangle ,$ $\left\vert 010\right\rangle ,$ $\left\vert
011\right\rangle ,$ $\left\vert 110\right\rangle $ and $\left\vert
111\right\rangle $. We define the operators used by Alice for measurement as%
\begin{eqnarray}
P^{k} &=&\$_{000}^{k}\pi _{000}+\$_{001}^{k}\pi _{001}+\$_{110}^{k}\pi
_{110}+\$_{010}^{k}\pi _{010}  \notag \\
&&+\$_{101}^{k}\pi _{101}+\$_{011}^{k}\pi _{011}+\$_{100}^{k}\pi
_{100}+\$_{111}^{k}\pi _{111}
\end{eqnarray}%
where $k=A,$ $B$ or $C$ and $\pi _{rst}$ can be written as a combination of
eight GHZ states%
\begin{eqnarray}
\pi _{000} &=&\left\vert \psi _{000}\right\rangle \left\langle \psi
_{000}\right\vert ,\qquad \left\vert \psi _{000}\right\rangle =\frac{1}{%
\sqrt{2}}(\left\vert 000\right\rangle +i\left\vert 111\right\rangle )  \notag
\\
\pi _{111} &=&\left\vert \psi _{111}\right\rangle \left\langle \psi
_{111}\right\vert ,\qquad \left\vert \psi _{111}\right\rangle =\frac{1}{%
\sqrt{2}}(\left\vert 111\right\rangle +i\left\vert 000\right\rangle )  \notag
\\
\pi _{001} &=&\left\vert \psi _{001}\right\rangle \left\langle \psi
_{001}\right\vert ,\qquad \left\vert \psi _{001}\right\rangle =\frac{1}{%
\sqrt{2}}(\left\vert 001\right\rangle +i\left\vert 110\right\rangle )  \notag
\\
\pi _{110} &=&\left\vert \psi _{110}\right\rangle \left\langle \psi
_{110}\right\vert ,\qquad \left\vert \psi _{110}\right\rangle =\frac{1}{%
\sqrt{2}}(\left\vert 110\right\rangle +i\left\vert 001\right\rangle )  \notag
\\
\pi _{010} &=&\left\vert \psi _{010}\right\rangle \left\langle \psi
_{010}\right\vert ,\qquad \left\vert \psi _{010}\right\rangle =\frac{1}{%
\sqrt{2}}(\left\vert 010\right\rangle -i\left\vert 101\right\rangle )  \notag
\\
\pi _{101} &=&\left\vert \psi _{101}\right\rangle \left\langle \psi
_{101}\right\vert ,\qquad \left\vert \psi _{101}\right\rangle =\frac{1}{%
\sqrt{2}}(\left\vert 101\right\rangle -i\left\vert 010\right\rangle )  \notag
\\
\pi _{011} &=&\left\vert \psi _{011}\right\rangle \left\langle \psi
_{011}\right\vert ,\qquad \left\vert \psi _{011}\right\rangle =\frac{1}{%
\sqrt{2}}(\left\vert 011\right\rangle -i\left\vert 100\right\rangle )  \notag
\\
\pi _{100} &=&\left\vert \psi _{100}\right\rangle \left\langle \psi
_{100}\right\vert ,\qquad \left\vert \psi _{100}\right\rangle =\frac{1}{%
\sqrt{2}}(\left\vert 100\right\rangle -i\left\vert 011\right\rangle )
\end{eqnarray}%
and $\$_{rst}^{k}$ are the real numbers as selected by the three parties
with mutual understanding.\emph{\ }The result of measurements performed by
Alice can be obtained as\emph{\ }%
\begin{equation}
\$^{k}(\theta _{i},\alpha _{A},\beta _{A})=\text{Tr}(P^{k}\rho _{f})\text{,}
\end{equation}%
\emph{\ } where Tr represents the trace of a\emph{\ }matrix .

Let in the quantum line from Alice to Bob or Charlie, there is an
eavesdropper, Eve, who performs the measurement on the qubit. The action of
measurement made by Eve on the qubit can be modeled as the action of phase
damping channel \cite{Chuang}.\ The quantum state after measurement
transforms to\emph{\ }%
\begin{equation}
\rho =\overset{2}{\underset{i=0}{\sum }}A_{i\text{ }}\rho _{in}\text{ }%
A_{i}^{\dagger }
\end{equation}%
where $A_{0}=\sqrt{p}\left\vert 0\right\rangle \left\langle 0\right\vert ,$ $%
A_{1}=\sqrt{p}\left\vert 1\right\rangle \left\langle 1\right\vert $ and $%
A_{2}=\sqrt{1-p}\hat{I}$ are the Kraus operators. An extension to $N$ qubits
is achieved by applying the measurement to each qubit in turn resulting 
\begin{equation}
\rho \rightarrow \sum\limits_{i=0}^{2}A_{i_{1}}\otimes A_{i_{2}}....\otimes
A_{i_{N}}\rho A_{i_{N}}^{\dagger }......\otimes A_{i_{2}}^{\dagger }\otimes
A_{i_{1}}^{\dagger }
\end{equation}%
Using equations (1), (4)-(5) and (7)-(9), the result of measurements
performed by Alice can be recorded as 
\begin{eqnarray}
&&\left. \Pi ^{k}(\theta _{i},\alpha _{A},\beta _{A})=\right.  \notag \\
&&\frac{c_{A}c_{B}c_{C}}{2}[(\$_{000}^{k}+\$_{111}^{k})+(\$_{000}^{k}-%
\$_{111}^{k})\mu _{p}\cos 2(\alpha _{A})]  \notag \\
&&+\frac{s_{A}s_{B}s_{C}}{2}[(\$_{000}^{k}+\$_{111}^{k})-(\$_{000}^{k}-%
\$_{111}^{k})\mu _{p}\cos 2(\beta _{A})]  \notag \\
&&+\frac{c_{A}c_{B}s_{C}}{2}[(\$_{001}^{k}+\$_{110}^{k})+(\$_{001}^{k}-%
\$_{110}^{k})\mu _{p}\cos 2(\alpha _{A})]  \notag \\
&&+\frac{s_{A}s_{B}c_{C}}{2}[(\$_{001}^{k}+\$_{110}^{k})-(\$_{001}^{k}-%
\$_{110}^{k})\mu _{p}\cos 2(\beta _{A})]  \notag \\
&&+\frac{s_{A}c_{B}c_{C}}{2}[(\$_{100}^{k}+\$_{011}^{k})+(\$_{100}^{k}-%
\$_{011}^{k})\mu _{p}\cos 2(\beta _{A})]  \notag \\
&&+\frac{c_{A}s_{B}s_{C}}{2}[(\$_{100}^{k}+\$_{011}^{k})-(\$_{100}^{k}-%
\$_{011}^{k})\mu _{p}\cos 2(\alpha _{A})]  \notag \\
&&+\frac{s_{A}c_{B}s_{C}}{2}[(\$_{101}^{k}+\$_{010}^{k})+(\$_{101}^{k}-%
\$_{010}^{k})\mu _{p}\cos 2(\beta _{A})]  \notag \\
&&+\frac{c_{A}s_{B}c_{C}}{2}[(\$_{101}^{k}+\$_{010}^{k})-(\$_{101}^{k}-%
\$_{010}^{k})\mu _{p}\cos 2(\alpha _{A})]
\end{eqnarray}%
where 
\begin{equation}
c_{i}=\cos ^{2}\frac{\theta _{i}}{2},\quad s_{i}=\sin ^{2}\frac{\theta _{i}}{%
2},\quad \mu _{p}=(1-p)
\end{equation}%
The elements of the decoding matrix for Bob and Charlie can be found by
putting the appropriate values for $\$_{rst}^{k}$ in the equation (10). Let
Alice, Bob and Charlie agree that $\$_{000}^{A}=\$_{000}^{B}=\$_{000}^{C}=3,$
$\$_{001}^{A}=\$_{001}^{B}=2,$ $\$_{001}^{C}=5,\
\$_{100}^{A}=5,\$_{100}^{B}=\$_{100}^{C}=2,$ $\$_{101}^{A}=\$_{101}^{C}=4,%
\$_{101}^{B}=0,$ $\$_{010}^{A}=\$_{010}^{C}=2,\$_{010}^{B}=5,$ $%
\$_{011}^{A}=0,\$_{011}^{B}=\$_{011}^{C}=4,$ $\$_{110}^{A}=\$_{110}^{B}=4,%
\$_{110}^{C}=0$ and $\$_{111}^{A}=\$_{111}^{B}=\$_{111}^{C}=1.$ This helps
Alice in establishing the decoding matrix.\emph{\ }For the case of no
eavesdropping i.e. $p=0$\ in equation (10), the decoding matrix becomes as
given in table 1.

Bob and Charlie, after applying one of their local unitary operators (known
to Alice) on their respective qubit, send them back to Alice, who first
applies her local unitary operators and then calculates the expectation
values of the coding operators. Then she compares the measured value with
the elements of the decoding matrix already present in her library
constructed by her simulation (table 1). Since she is well aware of her own
actions (unitary operations), therefore, she will have to compare only two
columns (one for Bob and the other for Charlie) of the decoding matrix
(table 1). By doing this she can easily find the unitary operators applied
by Bob and Charlie and hence she can find the corresponding secret key
element that they want to transmit her.\emph{\ }Repeating this process a
secret key is transferred from Bob and Charlie to Alice.

Whenever there is an eavesdropper, Eve, in the way and performs measurement
on the qubit, the decoding matrix in this case will change as given in
tables 2 and 3, representing the Charlie's actions $U_{C}\left( 0\right) $
and $U_{C}\left( \pi \right) $ respectively. It is easy to check from
elements of decoding matrices (tables 1, 2 and 3) that the matrix elements
are different from each other for the entire range of $p$ from $0$ to $1,$
for any particular choice of symbol to be transmitted from Bob or Charlie to
Alice, for example, $m_{1}$. In other words none of the elements of the
decoding matrix is repeated for any value of $p$ ranging from $0$ to $1$.
Hence, Eve can be detected very easily since Alice is well aware of her own
action. If the users of communication find the presence of Eve (from the
disturbance of decoding matrix elements), they\ will abort communication. By
repeating this process for several times, a string of bits can be
transferred from Bob and Alice to Alice which acts as a secret key for
communication.\emph{\ }In case of intercept/re-send attack, if Eve succeeds
in finding the qubit then the correlation between Alice and Bob or Alice and
Charlie will break and the elements of decoding matrix will change, giving
an indication of eavesdropping (which can be seen from tables 2 and 3). Then
the Alice will announce the abortion of communication to Bob and Charlie on
a classical channel.

\section{Security analysis}

As it is clear that the decoding\ matrix (table 1) is different from
decoding \ matrix (tables 2 and 3) for all values of $p>0$. Whenever for any
action of Bob or Charlie,\ if Alice finds the measured elements of the
decoding matrix different from the elements that she already has in her
library (table 1), then she would be able to detect the presence of Eve. One
of the most common eavesdropping strategy is the catch-and-resend attack. In
this attack if Eve succeeds in finding the bit, she can re-sends a similar
bit to Alice. But in our case if it so happens then the correlation between
Alice and Bob or Alice and Charlie will break and the elements of the
decoding matrix will change that reveals eavesdropping. For example, Bob
applies unitary operator $U_{B}\left( 0\right) $\ on his qubit and sends it
back to Alice, in the quantum line, Eve performs measurement on the qubit
and gets either $0$\ or $1$. On the bases of her measurement result, Eve
sends $\left\vert 0\right\rangle $\ or $\left\vert 1\right\rangle $\ to
Alice. If Alice applies $U_{A}\left( 0,0,0\right) $\ before measurement then
the final state received by her would be either $\left\vert 000\right\rangle 
$\ or $\left\vert 111\right\rangle $\ with equal probability. The
probability of getting the decoding matrix element $\left( \alpha _{i},\beta
_{i},\gamma _{i}\right) ,$ can be found for $n$ copies to be transmitted and
interrupted $i$ of them by Alice, using binomial distribution as 
\begin{equation}
P_{i}=\frac{1}{2^{n}}\binom{n}{i}
\end{equation}%
In this case the decoding matrix element can be obtained as 
\begin{equation}
\left( \alpha _{i},\beta _{i},\gamma _{i}\right) =\left( \frac{3(n-1)+5i}{n},%
\frac{3(n-1)+2i}{n},\frac{3(n-1)+2i}{n}\right)
\end{equation}%
Then finding the quantity 
\begin{equation}
f(n)=\left( \sum\limits_{i=0}^{n}P_{i}\alpha
_{i},\sum\limits_{i=0}^{n}P_{i}\beta _{i},\sum\limits_{i=0}^{n}P_{i}\gamma
_{i}\right) ,
\end{equation}%
we get $\left( 4,\frac{5}{2},\frac{5}{2}\right) $ which is independent of
the number of copies $n$. In addition, this is not an element of the
decoding matrix (table 1) for the corresponding action of Alice $U_{A}\left(
0,0,0\right) $ and hence Eve can be detected easily. Now we find out that
how many number of copies of input state, Alice requires for the detection
of Eve. The answer to this question can be given with the help of variance.
Since for the decoding matrix elements $\left( \alpha =3,\beta =3,\gamma
=3\right) $\ and $\left( \delta =5,\eta =2,\lambda =2\right) $\ (as seen
from table 1), the variance with respect to the number of copies $n$ varies
as 
\begin{equation*}
\left( \Delta _{1},\Delta _{2},\Delta _{3}\right) =\left( \frac{\alpha
-\delta }{2\sqrt{n}},\frac{\beta -\eta }{2\sqrt{n}},\frac{\gamma -\lambda }{2%
\sqrt{n}}\right) =\left( \frac{1}{\sqrt{n}},\frac{1}{2\sqrt{n}},\frac{1}{2%
\sqrt{n}}\right)
\end{equation*}%
Therefore for this case nine to ten copies are sufficient for Eve's
detection.

\section{Conclusions}

We devise a multiparty quantum cryptographic protocol using tripartite
entangled GHZ states. This protocol is efficient since two classical bits
can be transferred per entangled pair of qubits. In addition, in this
protocol same symbol can be used for key distribution and Eve's detection.
Unitary operators applied by Bob and Charlie on their part of tripartite
entangled state encodes a classical symbol that can be decoded at receiver's
end with the help of a decoding matrix. Eve's presence can be detected by
the disturbance of the decoding matrix. Furthermore, our protocol is secure
against intercept-resend attacks.

\begin{figure}[tbp]
\begin{center}
\vspace{-2cm} \includegraphics[scale=0.6]{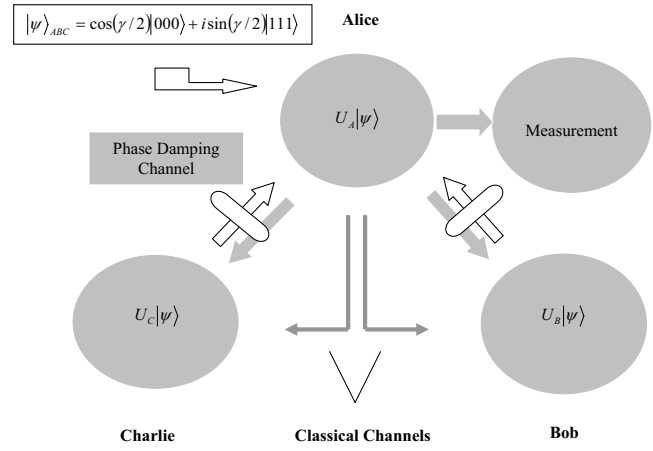} \\[0pt]
\end{center}
\caption{Schematic diagram of the model.}
\end{figure}
\newpage

\begin{table}[tbh]
\caption{The decoding matrix obtained as a result of Alice's simulations and
in the absence of Eve, i.e. $p=0$, where first number in the parenthesis
corresponds to Alice, the second number corresponds to Bob and the third
number corresponds to Charlie respectively.}
\label{di-fit}%
\begin{tabular}{l|l|l|l|l|}
\cline{2-5}
& \multicolumn{2}{|l|}{\ \ \ \ \ \ \ \ \ \ \ \ \ \ \ $%
\begin{tabular}{l}
$U_{C}(0)\rightarrow m_{3}$%
\end{tabular}%
$\ \ \ \ \ \ \ \ \ \ \ } & \multicolumn{2}{|l|}{\ \ \ \ \ \ \ \ \ \ \ \ \ \
\ $%
\begin{tabular}{l}
$U_{C}(\pi )\rightarrow m_{4}$%
\end{tabular}%
$\ \ \ \ \ \ } \\ 
& \multicolumn{2}{|l}{} & \multicolumn{2}{|l|}{} \\ \hline
\multicolumn{1}{|l|}{$%
\begin{tabular}{l}
Alice's Operation%
\end{tabular}%
$} & $%
\begin{tabular}{l}
$U_{B}(0)\rightarrow m_{1}$%
\end{tabular}%
$ & $%
\begin{tabular}{l}
$U_{B}(\pi )\rightarrow m_{2}$%
\end{tabular}%
$ & $%
\begin{tabular}{l}
$U_{B}(0)\rightarrow m_{1}$%
\end{tabular}%
$ & $%
\begin{tabular}{l}
$U_{B}(\pi )\rightarrow m_{2}$%
\end{tabular}%
$ \\ 
\multicolumn{1}{|l|}{} &  &  &  &  \\ \hline
\multicolumn{1}{|l|}{\ \ 
\begin{tabular}{l}
\\ 
\ \ $\ U_{A}(0,0,0)$\ \ \  \\ 
\\ 
\ \ $\ U_{A}(\pi ,\pi ,\pi )$ \\ 
\end{tabular}%
} & \ \ \ 
\begin{tabular}{l}
\\ 
(3,3,3) \\ 
\\ 
(5,2,2) \\ 
\end{tabular}
& \ \ \ \ 
\begin{tabular}{l}
\\ 
(2,5,2) \\ 
\\ 
(4,4,0) \\ 
\end{tabular}
& \ \ \ \ 
\begin{tabular}{l}
\\ 
(2,2,5) \\ 
\\ 
(4,0,4) \\ 
\end{tabular}
& \ \ \ \ 
\begin{tabular}{l}
\\ 
(0,4,4) \\ 
\\ 
(1,1,1) \\ 
\end{tabular}
\\ 
\multicolumn{1}{|l|}{} &  &  &  &  \\ \hline
\end{tabular}%
\end{table}
\newpage

\begin{table}[tbh]
\caption{The decoding matrix obtained as a result of measurements performed
by Alice (for Charlie's action $U_{C}\left( 0\right) $) in the presence of
Eve.}%
\begin{tabular}{l|l|l|}
\cline{2-3}
& \multicolumn{2}{|l|}{\ \ \ \ \ \ \ \ \ \ \ \ \ \ \ \ \ \ \ \ \ \ \ \ \ \ \
\ \ \ \ \ \ \ \ \ \ \ \ \ \ \ \ \ \ \ \ \ \ \ \ \ \ \ \ \ \ \ \ \ \ $%
\begin{tabular}{l}
$U_{C}(0)\rightarrow m_{3}$%
\end{tabular}%
$\ \ \ \ \ \ \ \ \ \ \ \ \ \ \ \ \ \ \ \ \ \ \ \ \ \ \ \ \ \ \ \ \ \ \ } \\ 
& \multicolumn{2}{|l|}{} \\ \hline
\multicolumn{1}{|l|}{$%
\begin{tabular}{l}
Alice's Operation%
\end{tabular}%
$} & \ \ $\ \ \ \ \ \ \ \ \ \ \ \ \ \ \ \ \ 
\begin{tabular}{l}
$U_{B}(0)\rightarrow m_{1}$%
\end{tabular}%
$\ \ \ \  & \ \ $\ \ \ \ \ \ \ \ \ \ \ \ \ \ \ \ \ \ \ 
\begin{tabular}{l}
$U_{B}(\pi )\rightarrow m_{2}$%
\end{tabular}%
$\ \ \ \ \ \  \\ 
\multicolumn{1}{|l|}{} &  &  \\ \hline
\multicolumn{1}{|l|}{\ \ 
\begin{tabular}{l}
\\ 
\ \ $\ U_{A}(0,0,0)$\ \ \  \\ 
\\ 
\ \ $\ U_{A}(\pi ,\pi ,\pi )$ \\ 
\end{tabular}%
} & \ \ \ 
\begin{tabular}{l}
\\ 
$(2+(1-p),2+(1-p),2+(1-p))$ \\ 
\\ 
$(\frac{5}{2}+\frac{5}{2}(1-p),3-(1-p),3-(1-p))$ \\ 
\end{tabular}
& \ \ \ \ 
\begin{tabular}{l}
\\ 
$(3-(1-p),\frac{5}{2}+\frac{5}{2}(1-p),3-(1-p))$ \\ 
\\ 
$(3+(1-p),3+(1-p),\frac{5}{2}-\frac{5}{2}(1-p))$ \\ 
\end{tabular}
\\ 
\multicolumn{1}{|l|}{} &  &  \\ \hline
\end{tabular}%
\end{table}
\newpage

\begin{table}[tbh]
\caption{The decoding matrix obtained as a result of measurements performed
by Alice (for Charlie's action $U_{C}\left( \protect\pi \right) $) in the
presence of Eve.}%
\begin{tabular}{l|l|l|}
\cline{2-3}
& \multicolumn{2}{|l|}{\ \ \ \ \ \ \ \ \ \ \ \ \ \ \ \ \ \ \ \ \ \ \ \ \ \ \
\ \ \ \ \ \ \ \ \ \ \ \ \ \ \ \ \ \ \ \ \ \ \ \ \ \ \ \ \ \ \ \ \ \ $%
\begin{tabular}{l}
$U_{C}(\pi )\rightarrow m_{4}$%
\end{tabular}%
$\ \ \ \ \ \ \ \ \ \ \ \ \ \ \ \ \ \ \ \ \ \ \ \ \ \ \ \ \ \ \ \ \ \ \ } \\ 
& \multicolumn{2}{|l|}{} \\ \hline
\multicolumn{1}{|l|}{$%
\begin{tabular}{l}
Alice's Operation%
\end{tabular}%
$} & \ \ $\ \ \ \ \ \ \ \ \ \ \ \ \ \ \ \ \ 
\begin{tabular}{l}
$U_{B}(0)\rightarrow m_{1}$%
\end{tabular}%
$\ \ \ \  & \ \ $\ \ \ \ \ \ \ \ \ \ \ \ \ \ \ \ \ \ \ 
\begin{tabular}{l}
$U_{B}(\pi )\rightarrow m_{2}$%
\end{tabular}%
$\ \ \ \ \ \  \\ 
\multicolumn{1}{|l|}{} &  &  \\ \hline
\multicolumn{1}{|l|}{\ \ 
\begin{tabular}{l}
\\ 
\ \ $\ U_{A}(0,0,0)$\ \ \  \\ 
\\ 
\ \ $\ U_{A}(\pi ,\pi ,\pi )$ \\ 
\end{tabular}%
} & \ \ \ 
\begin{tabular}{l}
\\ 
$(3-(1-p),3-(1-p),\frac{5}{2}+\frac{5}{2}(1-p))$ \\ 
\\ 
$(3+(1-p),\frac{5}{2}-\frac{5}{2}(1-p),3+(1-p))$ \\ 
\end{tabular}
& \ \ \ \ 
\begin{tabular}{l}
\\ 
$(\frac{5}{2}-\frac{5}{2}(1-p),3+(1-p),3+(1-p))$ \\ 
\\ 
$(2-(1-p),2-(1-p),2-(1-p))$ \\ 
\end{tabular}
\\ 
\multicolumn{1}{|l|}{} &  &  \\ \hline
\end{tabular}%
\end{table}

\end{document}